\documentclass[prb,reprint,nofootinbib,longbibliography,superscriptaddress]{revtex4-1}
\bibpunct{[}{]}{;}{n}{}{}

\pdfoutput=1
\usepackage[caption=false]{subfig}

\usepackage{graphicx}
\usepackage{comment}
\usepackage{amsmath}
\usepackage{amssymb}
\usepackage{amsfonts}
\usepackage{bbm}
\usepackage{braket}
\usepackage{verbatim}
\usepackage[colorlinks, urlcolor=blue]{hyperref}
\usepackage{ulem}
\usepackage{color, colortbl}
\usepackage{multirow}
\usepackage{physics}
\definecolor{LightCyan}{rgb}{1,0.5,0.5}
\hypersetup{
  colorlinks   = true, 
  urlcolor     = blue, 
  linkcolor    = blue, 
  citecolor   = blue 
}

\usepackage{cancel}
\usepackage{color}
\usepackage{tikz}

\newcommand{\be}{\begin{equation}}
\newcommand{\ee}{\end{equation}}

\definecolor{mscolor}{rgb}{0,0.5,0.5}

\begin{document}

	\title{Sampling Frequency Thresholds for Quantum Advantage of Quantum Approximate Optimization Algorithm}

\author{Danylo Lykov}
\email{dlykov@anl.gov}
\affiliation
{Computational Science Division, Argonne National Laboratory, 9700 S. Cass Ave., Lemont, IL 60439, USA}
\affiliation{Department of Computer Science, The University of Chicago, Chicago, IL 60637, USA}

\author{Jonathan Wurtz}
\affiliation{Department of Physics and Astronomy, Tufts University, Medford, MA 02155, USA}
\affiliation{QuEra Computing Inc., Boston, MA 02135, USA}

\author{Cody Poole}
\affiliation
{Department of Physics, University of Wisconsin - Madison, Madison, WI 53706, USA}

\author{Mark Saffman}
\affiliation
{Department of Physics, University of Wisconsin - Madison, Madison, WI 53706, USA}
\affiliation{ColdQuanta, Inc., 612 W. Main St., Madison, WI 53703, USA}

\author{Tom Noel}
\affiliation
{ColdQuanta, Inc., 3030 Sterling Circle, Boulder, CO 80301, USA}

\author{Yuri Alexeev}
\affiliation
{Computational Science Division, Argonne National Laboratory, 9700 S. Cass Ave., Lemont, IL 60439, USA}


	
	\begin{abstract}
 We compare the performance of the Quantum Approximate Optimization Algorithm (QAOA) with state-of-the-art classical solvers Gurobi and MQLib to solve the MaxCut problem on 3-regular graphs. We identify minimum noiseless sampling frequency and depth $p$ required for a quantum device to outperform classical algorithms. There is potential for quantum advantage on hundreds of qubits and moderate depth with a sampling frequency of ~10kHz. We observe, however, that classical heuristic solvers are capable of producing high-quality approximate solutions in linear time complexity.
In order to match this quality for large graph sizes $N$, a quantum device must support depth $p>11$.
Additionally, multi-shot QAOA is not efficient on large graphs, indicating that QAOA $p\leq11$ does not scale with~$N$. These results limit achieving quantum advantage for QAOA MaxCut on 3-regular graphs.
Other problems, such as different graphs, weighted MaxCut, and 3-SAT, may be better suited for achieving quantum advantage on near-term quantum devices.

	\end{abstract}

	\maketitle

\section{Introduction}

Quantum computing promises enormous computational powers that can far outperform any classical computational capabilities~\cite{alexeev2021quantum}. In particular, certain problems can be solved much faster compared with classical computing, as demonstrated experimentally by Google for the task of sampling from a quantum state~\cite{arute2019quantum}.
Thus,  an important milestone \cite{arute2019quantum} in quantum technology, so-called `quantum supremacy', was achieved as defined by Preskill~\cite{preskill2012quantum}. 

The next milestone, `quantum advantage', where quantum devices  solve useful problems faster than classical hardware, is more elusive and has arguably not  yet been demonstrated. However, a recent study suggests a possibility of achieving a quantum advantage in runtime over specialized state-of-the-art heuristic algorithms to solve the Maximum Independent Set problem using Rydberg atom arrays~\cite{ebadi2022}.
Common classical solutions to several potential applications for near-future quantum computing are heuristic and do not have performance bounds. Thus,  proving the advantage of quantum computers is far more challenging~\cite{Guerreschi2019,Zhou2020,Serret2020}.
Providing an estimate of how quantum advantage over these classical solvers can be achieved is important for the community and is the subject of this paper.

\begin{figure}[h!t] 

    \centering
    \includegraphics[width=\linewidth]{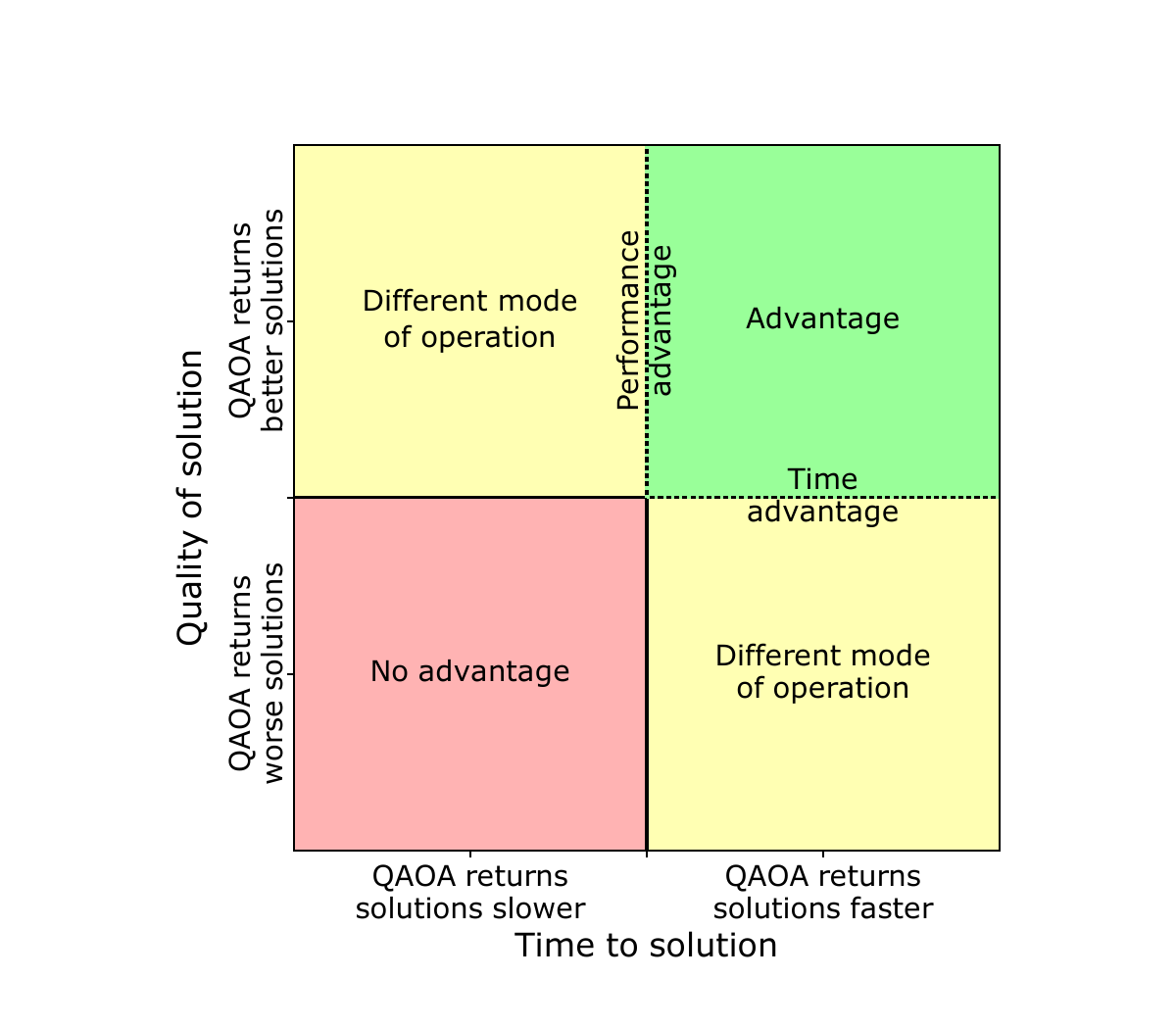}
    \caption{Locus of quantum advantage over classical algorithms. A particular classical algorithm may return some solution to some ensemble of problems in time $T_C$ (horizontal axis) with some quality $C_C$ (vertical axis). Similarly, a quantum algorithm may return a different solution sampled in time $T_Q$, which may be faster (right) or slower (left) than classical, with a better (top) or worse (bottom) quality than classical. If QAOA returns better solutions faster than the classical, then there is clear advantage (top right), and conversely  no advantage for worse solutions slower than the classical (bottom left).}
    \label{fig:comparison_map}
\end{figure}

Most of the useful quantum algorithms require large fault-tolerant quantum computers, which remain far in the future. In the near future, however, we can expect to have noisy intermediate-scale quantum (NISQ) devices~\cite{Preskill2018}. 
In this context variational quantum algorithms (VQAs) show the most promise~\cite{VQA_overview} for the NISQ era, such as the variational quantum eigensolver (VQE)~\cite{peruzzo2014variational} and the Quantum Approximate Optimization Algorithm~(QAOA)~\cite{farhi2014quantum}.
Researchers have shown remarkable interest in QAOA because it can be used to obtain approximate (i.e., valid but not optimal) solutions to a wide range of useful combinatorial optimization problems~\cite{ebadi2022, farhi2020quantum, Chatterjee2021}.

In opposition, powerful classical approximate and exact solvers have been developed to find good approximate solutions to combinatorial optimization problems. For example, a recent work by Guerreschi and Matsuura~\cite{Guerreschi2019} compares the time to solution of QAOA vs.~the classical combinatorial optimization suite AKMAXSAT. The classical optimizer takes exponential time with a small prefactor, which leads to the conclusion that QAOA needs hundreds of qubits to be faster than classical. This analysis requires the classical optimizer to find an exact solution, while QAOA  yields only approximate solutions. However, modern classical heuristic algorithms are able to return an approximate solution on demand.
Allowing for worse-quality solutions makes these solvers extremely fast (on the order of milliseconds), suggesting that QAOA must also be fast to remain competitive.
A valid comparison should consider both solution quality and time.

In this way, the locus of quantum advantage has two axes, as shown in Fig.~\ref{fig:comparison_map}: to reach advantage, a quantum algorithm must be both faster and return better solutions than a competing classical algorithm (green, top right). If the quantum version is slower and returns worse solutions (red, bottom left) there is clearly no advantage. However,  two more regions are shown in the figure. If the QAOA returns better solutions more slowly than a classical algorithm (yellow, top left), then we can increase the running time for the classical version. It can try again and improve its solution with more time. This is a crucial mode  to consider when assessing advantage: heuristic algorithms may always outperform quantum algorithms if quantum time to solution is slow.
Alternatively, QAOA may return worse solutions faster (yellow, bottom right), which may be useful for time-sensitive applications.
In the same way, we may stop the classical algorithm earlier, and the classical solutions will become worse.

One must keep in mind that the reason for using a quantum algorithm is the scaling of its time to solution with the problem size $N$. 
Therefore, a strong quantum advantage claim should demonstrate the superior performance of a quantum algorithm in the large-$N$ limit.

This paper focuses on the MaxCut combinatorial optimization problem on 3-regular graphs for various problem size $N$. MaxCut is a popular benchmarking problem for QAOA because of its simplicity and straightforward implementation.
We propose a fast fixed-angle approach to running QAOA that speeds up QAOA while preserving solution quality compared with slower conventional approaches.
We evaluate the expectation value of noiseless QAOA solution quality using tensor network simulations on classical hardware. We then find the time required for classical solvers to match this expected QAOA solution quality.
Surprisingly, we observe that even for the smallest possible time, the classical solution quality is above our QAOA solution quality for $p=11$, our largest $p$ with known performance.
Therefore, we compensate for this difference in quality by using multishot QAOA and find the number of samples $K$ required to match the classical solution quality.
$K$ allows us to characterize quantum device parameters, such as sampling frequency, required for the quantum algorithm to match the classical solution quality.

\section{Results and discussion}

This section will outline the results and comparison between classical optimizers and QAOA. This has two halves: Sec.~\ref{sec:QAOA_results} outlines the results of the quantum algorithm, and Sec.~\ref{sec:classical_results} outlines the results of the classical competition.

\label{sec:results}

\subsection{Expected QAOA solution quality}\label{sec:QAOA_results}

The first algorithm is the quantum approximate optimization algorithm (QAOA), which uses a particular ansatz to generate approximate solutions through measurement.  We evaluate QAOA for two specific modes. The first is single shot fixed angle QAOA, where a single solution is generated. This has the the benefit of being very fast. The second generalization is multi-shot fixed angle QAOA, where many solutions are generated, and the best is kept. This has the benefit that the solution may be improved with increased run time.

In Section \ref{sec:QAOAperformance} we find that one can put limits on the QAOA MaxCut performance even when the exact structure of a 3-regular graph is unknown using fixed angles. We have shown that for large $N$ the average cut fraction for QAOA solutions on 3-regular graphs converges to a fixed value $f_{\text {tree}}$. If memory limitations permit, we evaluate these values numerically using tensor network simulations. This gives us the average QAOA performance for any large $N$ and $p\leq11$.  To further strengthen the study of QAOA performance estimations, we verify that for the small $N$, the performance is close to the same value $f_{\text{tree}}$. We are able to numerically verify that for $p\leq 4$ and small $N$ the typical cut fraction is close to $f_{\text {tree}}$, as shown on Fig.~\ref{fig:bounds_p1}.

Combining the large-$N$ theoretical analysis and small-$N$ heuristic evidence, we are able to predict the average performance of QAOA on 3-regular graphs for $p\leq11$. We note that today's hardware can run QAOA up to $p\leq4$ \cite{ebadi2022} and that for larger depths the hardware noise prevents achieving better QAOA performance. Therefore, the $p\leq11$ constraint is not an important limitation for our analysis.

\begin{figure}
    \centering
    \includegraphics[width=\linewidth]{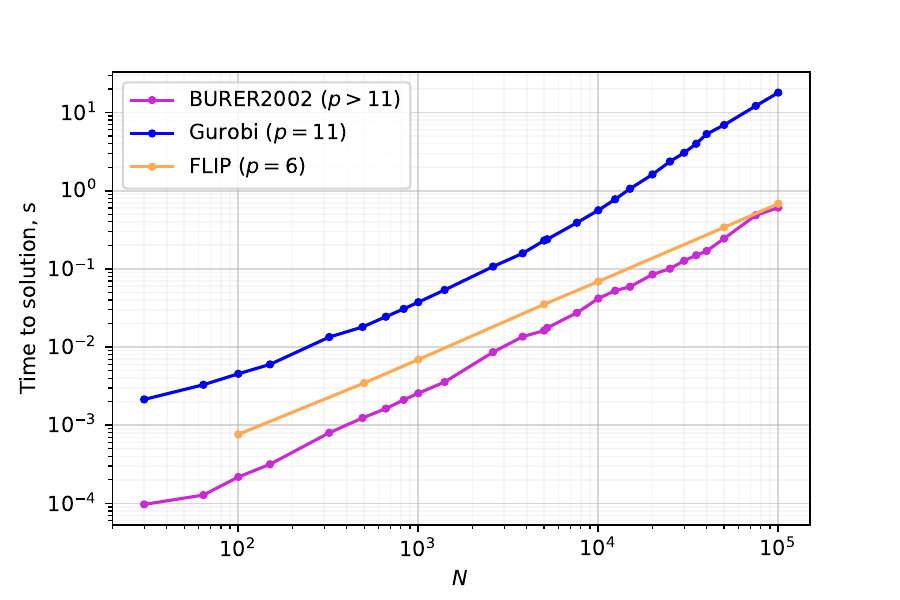}
    \caption{
    Time required for a single-shot QAOA to match classical MaxCut algorithms.
    The blue line shows time for comparing with the Gurobi solver and using $p=11$;  the yellow line shows comparison with the FLIP algorithm and $p=6$.
    Each quantum device that runs MaxCut QAOA can be represented on this plot as a point, where the x-axis is the number of qubits and the y-axis is the time to solution. 
    For any QAOA depth $p$, the quantum device should return at least one bitstring faster than the Y-value on this plot.
    }
    \label{fig:adv_freq}
\end{figure}

\subsection{Classical solution quality and time to solution}\label{sec:classical_results}

\begin{figure}
    \includegraphics[width=\linewidth]{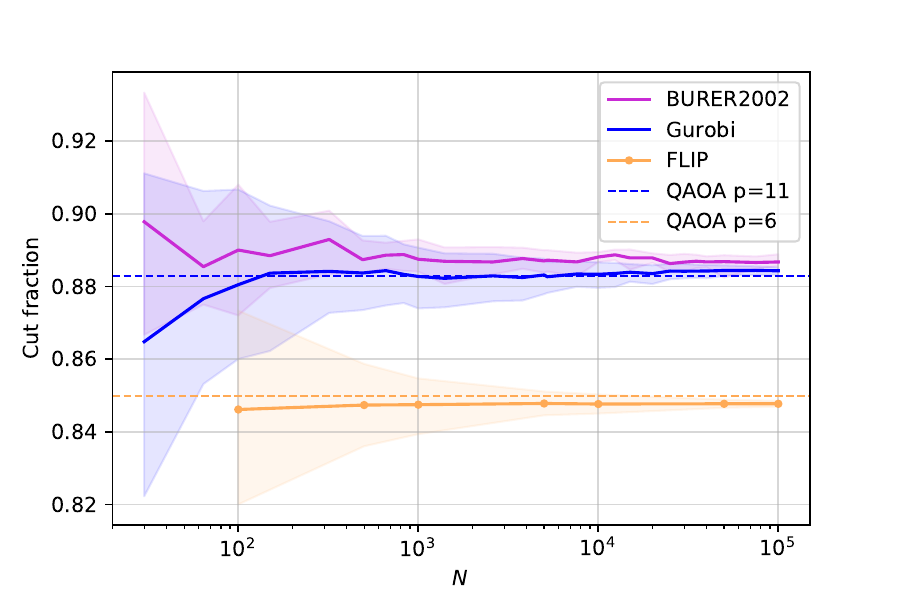}
    \caption{
    Zero-time performance for graphs of different size $N$.
    The Y-value is the cut fraction obtained by running corresponding algorithms for minimum possible time. This corresponds to the Y-value of the star marker in Fig.~\ref{fig:timebounds}.
    Dashed lines show the expected QAOA performance for $p=11$ (blue) and $p=6$ (yellow). QAOA can outperform the FLIP algorithm at depth $p>6$, while for Gurobi it needs $p>11$.
    Note that in order to claim advantage, QAOA has to provide the zero-time solutions in faster time than FLIP or Gurobi does. These times are shown on Fig.~\ref{fig:adv_freq}.
    }
    \label{fig:t0_cutf}
\end{figure}

The second ensemble of algorithms are classical heuristic or any-time algorithms. These algorithms have the property that they can be stopped mid-optimization and provide the best solution found so far. After a short time spent loading the instance, they find an initial `zero-time' guess. Then, they explore the solution space and find incramentally better solutions until stopping with the best solution after a generally exponential amount of time. We experimentally evaluate the performance of the classical solvers Gurobi, MQLib using BURER2002 heuristic, and FLIP in Sec.~\ref{sec:meth_classical}. We observe that the zero-time performance, which is the quality of the fastest classical solution, is above the expected quality of QAOA $p=11$, as shown in Fig.~\ref{fig:t0_cutf}. The time to first solution scales almost linearly with size, as shown in Fig.~\ref{fig:adv_freq}. To compete with classical solvers, QAOA has to return better solutions faster.

\subsection{Multi-shot QAOA}

To improve the performance of QAOA, one can sample many bitstrings and then take the best one. This approach will  work only  if the  dispersion of the cut fraction distribution is large, however. For example, if the  dispersion is zero, measuring the ansatz state would return only bitstrings with a fixed cut value. By analyzing the correlations between the qubits in Section \ref{sec:QAOAperformance}, we show that the distribution of the cut fraction is a Gaussian with the  standard deviation on the order of $1/\sqrt N$. The expectation value of maximum of $K$ samples is proportional to the  standard deviation, as shown in Equation \ref{eq:multi-shot}. This equation determines the performance of multishot QAOA. In the large $N$ limit the  standard deviation is small, and one might need to measure more samples in order to match the classical performance.

If we have the mean performance of a classical algorithm, we can estimate the number of samples $K$ required for QAOA to match the classical performance. We denote the difference between classical and quantum expected cut fraction as $\Delta_p(t)$, which is a function of the running time of the classical algorithm. Moreover, it also depends on $p$, since $p$ determines QAOA expected performance. If $\Delta_p(t) < 0$, the performance of QAOA is better, and we  need only a $K=1$ sample. In order to provide an advantage, QAOA would have to measure this sample faster than the classical algorithm, as per Fig.~\ref{fig:comparison_map}. On the other hand, if $\Delta_p(t) > 0$, the classical expectation value is larger than the quantum one, and we have to perform multisample QAOA. We can find $K$ by inverting Equation \ref{eq:multi-shot}. In order to match the classical algorithm, a quantum device should be able to run these $K$ samples in no longer than $t$. We can therefore get the threshold sampling frequency.
\begin{equation}
   \nu_p(t) = \frac{K}{t} = \frac{1}{t}\exp \left({\frac{N}{2\gamma_p^2}\Delta_p(t) ^2} \right)
\end{equation}
The scaling of $\Delta_p(t)$ with $t$ is essential here since it determines at which point $t$ we will have the smallest sampling frequency for advantage. We find that for BURER2002, the value of $\Delta(t)$ is the lowest for the smallest possible $t=t_0$, which is when a classical algorithm can produce its first solution. To provide the lower bound for QAOA we consider $t_0$ as the most favourable point, since classical solution improves much faster with time than a multi-shot QAOA solution. This point is discussed in more detail in the Supplementary Methods.

\begin{figure}
    \centering
    \includegraphics[width=\linewidth]{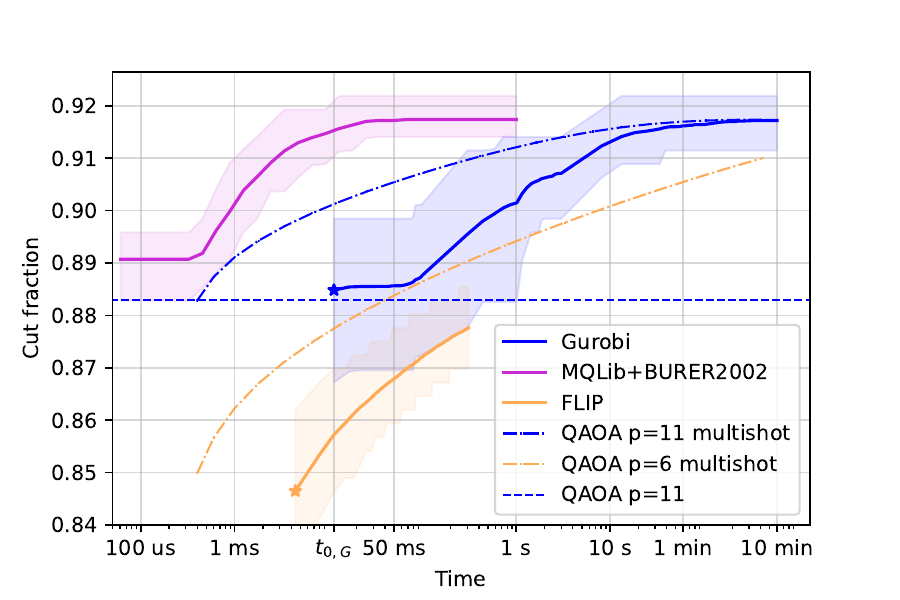}
    \caption{
    Evolution of cut fraction value in the process of running the classical algorithms solving 3-regular MaxCut with $N$=256. The shaded area shows 90-10 percentiles interval, 
    and the solid line shows the mean cut fraction over 100 graphs.
    The dashed lines show the expectation value of single-shot QAOA for $p=6, 11$, and
    the dash-dotted lines show the expected performance for multishot QAOA given a sampling rate of 5 kHz.
    Note that for this $N=256$ the multi-shot QAOA with $p=6$ can compete with Gurobi
    at 50 milliseconds. However, the slope of the multi-shot line will decrease for larger $N$, reducing the utility of the multi-shot QAOA. 
    }
    \label{fig:timebounds}
\end{figure}

Time $t_0$ is shown on Fig.~\ref{fig:adv_freq} for different classical algorithms. We note that in the figure
the time scales polynomially with the number of nodes $N$. Figure~\ref{fig:t0_cutf} shows the mean cut fraction for the same classical algorithms, as well as the expectation value of QAOA at $p=6, 11$. These two figures show that a simple linear-runtime FLIP algorithm is fast and gives a performance on par with $p=6$ QAOA. In this case $\Delta_6(t_0) < 0$, and we  need to sample only a single bitstring. To obtain the $p=6$ sampling frequency for advantage over the FLIP algorithm, one has to invert the time from Fig.~\ref{fig:adv_freq}. If the quantum device is not capable of running $p=6$ with little noise, the quantum computer will have to do multishot QAOA. Note that any classical prepossessing for QAOA will be at least linear in time since one must read the input and produce a quantum circuit. Therefore, for small $p<6$ QAOA will not give significant advantage: for any fast QAOA device one needs a fast classical computer; one might just run the classical FLIP algorithm on it.

\begin{figure}
    \centering
    \includegraphics[width=\linewidth]{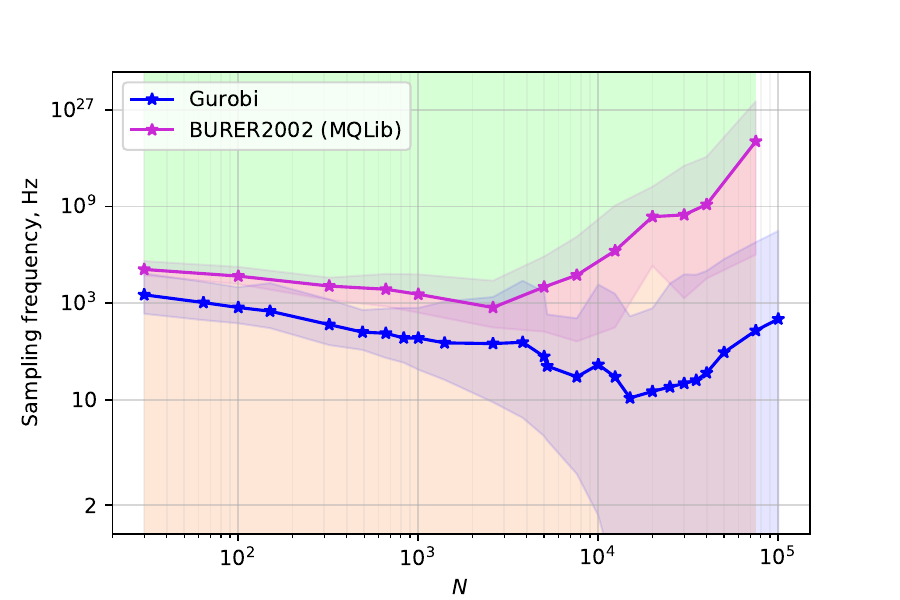}
    \caption{
    Sampling frequency required to achieve MaxCut advantage using QAOA $p=11$.
    The shaded area around the solid lines corresponds to 90-10 percentiles over 100 seeds for Gurobi and 20 seeds for BURER2002.
    The background shading represents comparison of a quantum computer with BURER2002 solver corresponding to modes in Fig.~\ref{fig:comparison_map}.
    Each quantum device can be represented on this plot as a point, where the x-axis is the number of qubits, and the y-axis is the time to solution. 
    Depending on the region where the point lands, there are different results of comparisons.
    QAOA becomes inefficient for large $N$, when sampling frequency starts to grow exponentially with $N$.
    }
    \label{fig:adv_rate}
\end{figure}

The Gurobi solver is able to achieve substantially better performance, and it slightly outperforms $p=11$ QAOA. Moreover, the BURER2002 algorithm demonstrates even better solution quality than does  Gurobi while being significantly faster. For both Gurobi and BURER2002, the $\Delta_{11}(t_0) > 0$, and we need to either perform multishot QAOA or increase $p$. Figure~\ref{fig:adv_rate} shows the advantage sampling frequency $\nu_{11}(t_0)$ for the Gurobi and BURER2002 algorithms; note that the vertical axis is doubly exponential.

The sampling frequency is a result of two factors that work in opposite directions. On the one hand, the time to solution for a classical algorithm grows with $N$, and hence $\nu$ drops. On the other hand, the  standard deviation of distribution vanishes as $1/\sqrt{N}$, and therefore the number of samples $K$ grows exponentially. There is an optimal size $N$ for which the sampling frequency is minimal.
This analysis shows that there is a possibility for advantage with multi-shot QAOA for moderate sizes of $N=100..10\,000$, for which a sampling frequency of $\approx10$kHz is required.
These frequencies are very sensitive to the difference in solution quality, and for $p\geq12$ a different presentation is needed, if one quantum sample is expected to give better than classical solution quality. This is discussed in more detail in Supplementary Methods.

For large $N$, as expected, we see a rapid growth of sampling frequency, which indicates that QAOA does not scale for larger graph sizes, unless we go to higher depth $p>11$. The color shading shows correspondence with Fig.~\ref{fig:comparison_map}. If the quantum device is able to run $p \geq 11$ and its sampling frequency and the number of qubits $N$ corresponds to the green area, we have a quantum advantage. Otherwise, the quantum device belongs to the red area, and there is no advantage.

It is important to note the effect of classical parallelization on our results. Despite giving more resources to the classical side, parallel computing is unlikely to help it. To understand this, one has to think on how parallelization would change the performance profile as shown on Figure~\ref{fig:timebounds}. The time to the first classical solution is usually bound from below by preparation tasks such as reading the graph, which are inherently serial. Thus, parallelization will not reduce $t_0$ and is in fact likely to increase it due to communication overhead. Instead, it will increase the slope of the solution quality curve, helping classical algorithms to compete in the convergence regime.

\subsection{Discussion}

As shown in Fig.~\ref{fig:comparison_map}, to achieve quantum advantage, QAOA must return better solutions faster than the competing classical algorithm. This puts stringent requirements on the speed of QAOA, which previously may have gone unevaluated. If QAOA returns a solution more slowly, the competing classical algorithm may `try again' to improve its solution, as is the case for anytime optimizers such as the Gurobi solver. The simplest way to improve the speed of QAOA is to reduce the number of queries to the quantum device, which we propose in our fixed-angle QAOA approach. This implementation forgoes the variational optimization step and uses solution concentration, reducing the number of samples to order 1 instead of order 100,000. Even with these improvements, however, the space of quantum advantage may be difficult to access.

Our work demonstrates that with a quantum computer of $\approx 100$ qubits, QAOA can be competitive with classical MaxCut solvers if the time to solution is shorter than 100~$\mu$s and the depth of the QAOA circuit is $p\geq6$. Note that this time to solution must include all parts of the computation, including state preparation, gate execution, and measurement. Depending on the parallelization of the architecture, there may be a quadratic time overhead.
However, the required speed of the quantum device grows with $N$ exponentially. Even if an experiment shows advantage for intermediate $N$ and $p\leq11$, the advantage will be lost on larger problems regardless of the quantum sampling rate. 
Thus, in  order  to be fully competitive with classical MaxCut solvers, quantum computers have to increase solution quality, for instance by using $p\geq12$.
Notably, $p=12$ is required but not sufficient for achieving advantage: the end goal is obtaining a cut fraction better than $\geq 0.885$ for large $N$, including overcoming other challenges of quantum devices such as noise.

These results lead us to  conclude that for 3-regular graphs (perhaps all regular graphs),  achieving quantum advantage on NISQ devices may be difficult. For example, the fidelity requirements to achieve quantum advantage are well above the characteristics of NISQ devices.

We note that  improved versions of QAOA exist, where the initial state is replaced with a preoptimized state \cite{warmstartQAOA} or the mixer operator is adapted to improve performance \cite{zhu2020adaptive, Govia2021}. One also can use information from classical solvers to generate a better ansatz state \cite{wurtz2021classically}. These algorithms have further potential to compete against classical MaxCut algorithms.
Also, more general problems, such as weighted MaxCut, maximum independent set, and 3-SAT, may be necessary in order to find problem instances suitable for achieving quantum advantage.

When comparing with classical algorithms, one must record the complete time to solution from the  circuit configuration to the measured state. This parameter may be used in the extension of the notion of quantum volume, which is customarily used for quantum device characterization.
Our work shows that QAOA MaxCut does not scale with graph size for at least up to $p\leq11$, thus putting quantum advantage for this problem away from the NISQ era.

\section{Methods}\label{sec:methodology}

Both classical solvers and QAOA return a bitstring as a solution to the MaxCut problem. To compare the algorithms,
we must decide on a metric to use to measure the quality of the solution. A common metric for QAOA and many classical algorithms is the approximation ratio, which is defined as the ratio of cut value (as defined in Eq.~(\ref{eq:maxcut_cost})) of the solution divided by the optimal (i.e., maximum possible) cut value for the given graph.
This metric is hard to evaluate heuristically for large $N$, since we do not know the optimal solution. We therefore use the cut fraction as the metric for solution quality, which is the cut value divided by the number of edges.

We analyze the algorithms on an ensemble of problem instances. Some instances may give advantage, while others may not. We 
therefore analyze ensemble advantage,
which compares the average solution quality over the ensemble.
The set of 3-regular graphs is extremely large for large graph size $N$, so for classical heuristic algorithms we evaluate the performance on a subset of graphs.
We then look at the mean of the cut fraction over the ensemble, which is the statistical approximation of the mean of the cut fraction over all 3-regular graphs.

\subsection{QAOA Methodology}

Usually  QAOA is thought of as a hybrid algorithm, where a quantum-classical outer loop optimizes the angles $\gamma,\beta$ through repeated query to the quantum device by a classical optimizer. Depending on the noise, this process may require hundreds or thousands of queries in order to find optimal angles, which slows the computation. To our knowledge,  no comprehensive work exists on exactly how many queries may be required to find such angles. It has been numerically observed \cite{Shaydulin_2019,Zhou2020}, however, that for small graph size $N=12$ and $p=4$, classical noise-free optimizers may find good angles in approximately $100$ steps, which can be larger for higher $N$ and $p$. Each step may need order $10^3$  bitstring queries to average out shot noise and find expectation values for an optimizer, and thus seeking global angles may require approximately $100\,000$ queries to the simulator.
The angles are then used for preparing an ansatz state, which is in turn measured (potentially multiple times) to obtain a solution.
Assuming a sampling rate of 1 kHz, this approach implies a QAOA solution of approximately  $100$ seconds.

 Recent results, however, suggest that angles may be precomputed on a classical device \cite{streif2019training} or transferred from other similar graphs \cite{galda2021transferability}.
 Further research analytically finds optimal angles for~$p\leq20$~and $d\to\infty$ 
 for all large-girth $d$-regular graphs, but does not give angles for finite~$d$~\cite{Basso2022}.
 Going a step further, a recent work finds that evaluating regular graphs at particular fixed angles has good performance on all problem instances \cite{Wurtz_guarantee}. These precomputed or fixed angles allow the outer loop to be bypassed, finding close to optimal results in a single shot. In this way, a $1000$ Hz QAOA solution can be found in  milliseconds, a speedup of several ordesr of magnitude.

For this reason we study the prospect for quantum advantage in the context of fixed-angle QAOA. For $d$-regular graphs, there exist particular fixed angles with universally good performance \cite{wurtz2021fixed}. 
Additionally, as will be shown in Section \ref{sec:single-shot}, one can reasonably expect that sampling a single bitstring from the fixed-angle QAOA will yield a solution with a cut fraction close to the expectation value.

The crucial property of the fixed-angle single-shot approach is that it is guaranteed to work for any graph size $N$.
On the other hand, angle optimisation could be less productive for large $N$, and the multiple-shot (measuring the QAOA ansatz multiple times) approach is less productive for large $N$, as shown in Section \ref{sec:multi-shot}.
Moreover, the quality of the solution scales with depth as $\sqrt p$ \cite{wurtz2021fixed}, which is faster than with the number of samples $\sqrt{\log K}$, instructing us to resort to multishot QAOA only if larger $p$ is unreachable.
Thus, the fixed-angle single-shot QAOA can robustly speed up finding a good approximate solution from the order of seconds to milliseconds, a necessity for advantage over state-of-the-art anytime heuristic classical solvers, which can get good or exact solutions in approximately milliseconds. Crucially, single-shot QAOA quality of solution can be maintained for all sizes $N$ at fixed depth $p$, which can mean constant time scaling, for particularly capable quantum devices.

To simulate the expectation value of the cost function for 
QAOA, we employ a classical quantum circuit simulation algorithm QTensor \cite{lykov2021large, lykov_diagonal, LykovGPU}.
This algorithm is based on tensor network contraction and is described in more detail in Supplementary Methods.
Using this approach, one can simulate expectation values on a classical computer even for circuits with millions of qubits.

\subsection{Classical Solvers}
\label{sec:meth_classical}

Two main types of classical MaxCut algorithms exist: approximate algorithms and heuristic solvers.
Approximate algorithms guarantee a certain quality of solution for any problem instance. Such algorithms~\cite{Halperin_MaxCut,gw-algo} 
also provide polynomial-time scaling.
Heuristic solvers~\cite{gurobi, MQLib} are usually based on branch-and-bound methods~\cite{gurobi_mip} that use branch pruning and heuristic rules for  variable and value ordering. These heuristics are usually 
designed to run well on graphs that are common in practical use cases.
Heuristic solvers typically return better solutions than do approximate solvers,
but they provide no guarantee on the quality of the solution.

The comparison of QAOA with classical solvers thus requires making choices of measures that depend on the context of comparison. From a theory point of view, guaranteed performance is more important; in contrast, from an applied point of view, heuristic performance is the measure of choice.
A previous work \cite{Wurtz_guarantee} demonstrates that
QAOA provides better performance guarantees than does the Goemans--Williamson algorithm \cite{gw-algo}.
In this paper we  compare against heuristic algorithms since such a comparison is more 
relevant for real-world problems.
On the other hand, the performance of classical solvers reported in this paper
can depend on a particular problem instance.

We evaluate two classical algorithms using a single node of Argonne's Skylake testbed; the processor used is an Intel Xeon Platinum 8180M CPU @ 2.50 GHz with 768 GB of RAM.

The first algorithm we study is the Gurobi solver \cite{gurobi}, which is a combination of many heuristic algorithms.
We evaluate Gurobi with an improved configuration based on communication with 
Gurobi support~\footnote{\url{https://support.gurobi.com/hc/en-us/community/posts/4403570181137-Worse-performance-for-smaller-problem}}.
We use \texttt{Symmetry=0} and \texttt{PreQLinearize=2} in our improved configuration.
As further tweaks and hardware resources may increase the speed, the results here serve as a characteristic lower bound on Gurobi performance rather than a true guarantee. 
We run Gurobi on 100 random-regular graphs for each size $N$ and allow each optimization to run for 30 minutes.
During the algorithm runtime we collect information about the process, in particular the quality of the best-known solution.
In this way we obtain a performance profile 
of the algorithm that shows the relation between the solution quality and the running time. An example of such a performance profile for $N=256$ is shown in Fig. \ref{fig:timebounds}.
Gurobi was configured to use only a single CPU, to avoid interference in runtime between different Gurobi optimization runs for different problem instances. In order to speed up collection of the statistics, 
55 problem instances were executed in parallel.

The second algorithm is MQLib \cite{MQLib},
which is implemented in C++ and uses a variety of different heuristics
for solving MaxCut and QUBO problems.
We chose the BURER2002 heuristic since in our experiments it 
performs the best for MaxCut on random regular graphs.
Despite using a single thread, this algorithm is much faster than Gurobi; thus we run it for 1 second.
In the same way as with Gurobi, we collect the
performance profile of this algorithm.

While QAOA and Gurobi can be used as general-purpose combinatorial optimization algorithms, this algorithm is designed to
solve MaxCut problems only, and the heuristic was picked 
that demonstrated the best performance on the graphs we considered.
In this way we use Gurobi as a worst-case classical solver,
which is capable of solving the same problems as QAOA can.
Moreover, Gurobi is a well-established commercial tool that is widely used in industry. 
Note, however, that we use QAOA fixed angles that are optimized specifically for 3-regular graphs, and one can argue
that our fixed-angle QAOA is an algorithm designed for 3-regular MaxCut.
For this reason we also consider the best-case MQLib+BURER2002 classical algorithm, which is designed for MaxCut, and we choose the heuristic that performs best on 3-regular graphs.

\subsection{QAOA performance}\label{sec:QAOAperformance}

Two aspects are involved in  comparing the performance of algorithms, as outlined in Fig.~\ref{fig:comparison_map}: time to solution and quality of solution. In this section we evaluate the performance of single-shot fixed-angle QAOA. As discussed in the introduction, the time to solution is a crucial part and for QAOA is dependent on the initialization time and the number of rounds of sampling. Single-shot fixed-angle QAOA involves  only a single round of sampling, and so the time to solution can be extremely fast, with initialization time potentially becoming the limiting factor.
This initialization time is bound by the speed of classical computers, which perform calibration and device control. Naturally, if one is able to achieve greater initialization speed by using better classical computers, the same computers can be used to improve the speed of solving MaxCut classically. Therefore, it is also important to consider the time scaling of both quantum initialization and classical runtime.

The quality of the QAOA solution is the other part of performance. The discussion below evaluates this feature by using subgraph decompositions and QAOA typicality, including a justification of single shot sampling.

 QAOA is a variational ansatz algorithm structured to provide solutions to combinatorial optimization problems. The ansatz is constructed as $p$ repeated applications of an objective $\hat C$ and mixing $\hat B$ unitary:
\begin{equation}\label{eq:QAOA_ansatz}
|\gamma,\beta\rangle = e^{-i\beta_p \hat B}e^{-i\gamma_p \hat C}(\cdots)e^{-i\beta_1 \hat B}e^{-i\gamma_1 \hat C}|+\rangle ,
\end{equation}
where $\hat B$ is a sum over Pauli $X$ operators $\hat B = \sum_i^N\hat \sigma_x^i$. A common problem instance is MaxCut, which strives to bipartition the vertices of some graph $\mathcal G$ such that the maximum number of edges have vertices in opposite sets. Each such edge is considered to be cut by the bipartition. This may be captured in the objective function

\begin{equation}
\hat C = \frac{1}{2}\sum_{\langle ij\rangle \in \mathcal G}(1 - {\hat \sigma}_z^i {\hat \sigma}_z^j),
\label{eq:maxcut_cost}
\end{equation}
\newcommand{\gb}{{\vb*{\gamma}, \vb*{\beta}}}
\newcommand{\Z}{{\hat\sigma_z}}
whose eigenstates are bipartitions in the $Z$ basis, with eigenvalues that count the number of cut edges.
To get the solution to the optimization problem, one prepares the ansatz state $\ket \gb$ on a quantum device and then measures the state. The measured bitstring is the solution output from the algorithm.

While  QAOA is guaranteed to converge to the exact solution in the $p\to\infty$ limit in accordance with the adiabatic theorem \cite{farhi2014quantum,wurtz2021counterdiabaticity}, today's hardware is limited to low depths $p\sim 1$ to $5$, because of the noise and decoherence effects inherent to the NISQ era.

A useful tool for analyzing the performance of QAOA is  the fact that QAOA is local \cite{farhi2014quantum,farhi2020quantum}: the entanglement between any two qubits at a distance of $\geq2p$ steps from each other is strictly zero. For a similar reason, the expectation value of a particular edge $\langle ij\rangle$

\begin{equation}\label{eq:edge_expectation}
    f_{\langle ij\rangle} = \frac{1}{2}\langle \gb|1 - \hat \sigma_z^i\hat\sigma_z^j|\gb \rangle
\end{equation}
depends only on the structure of the graph within $p$ steps of edge $\langle ij\rangle$. Regular graphs have a finite number of such local structures (also known as subgraphs) \cite{Wurtz_guarantee}, and so the expectation value of the objective function can be rewritten as a sum over subgraphs

\begin{equation}
    \langle \hat C\rangle = \sum_{\text{subgraphs } \lambda}M_\lambda(\mathcal G) f_\lambda.
\end{equation}

Here, $\lambda$ indexes the different possible subgraphs of depth $p$ for a $d$ regular graph, $M_\lambda(\mathcal G)$ counts the number of each subgraph $\lambda$ for a particular graph $\mathcal G$, and $f_\lambda$ is the expectation value of the subgraph (e.g.,~Eq.~\eqref{eq:edge_expectation}). For example, if there are no cycles $\leq 2p+1$,  only one subgraph (the tree subgraph)  contributes to the sum.

With this tool we may ask and answer the following question: What is the typical performance of single-shot fixed-angle QAOA, evaluated over some ensemble of graphs? Here, performance is characterized as the typical (average) fraction of edges cut by a bitstring solution returned by a single sample of fixed-angle QAOA, averaged over all graphs in the particular ensemble. 

For our study we choose the ensemble of $3$-regular graphs on $N$ vertices. Different ensembles, characterized by different connectivity $d$ and size $N$, may have different QAOA performance \cite{herrman2021impact, shaydulin2021qaoakit}.

Using the structure of the random regular graphs, we can put bounds on the cut fraction by bounding the number of different subgraphs and evaluating the number of large cycles. 
These bounds become  tighter for $N\longrightarrow\infty$ and fixed $p$ since the majority of subgraphs become trees and 1-cycle graphs.
We describe this analysis in detail in Supplemental methods, which shows that the QAOA cut fraction will equal the expectation value on the tree subgraph, which may be used as a `with high probability' (WHP) proxy of performance. Furthermore, using a subgraph counting argument, we may count the number of tree subgraphs to find an upper and lower WHP bound on the cut fraction for smaller graphs. These bounds are shown as the boundaries of the red and green regions in Fig.~\ref{fig:bounds_p1}.

\subsection{QAOA Ensemble Estimates}\label{sec:ensemble_estimates}

A more straightforward but less rigorous characterization of QAOA performance is simply to evaluate fixed-angle QAOA on a subsample of graphs in the ensemble. The results of such an analysis require an assumption not on the particular combinatorial graph structure of ensembles but instead on the typicality of expectation values on subgraphs. This is an assumption on the structure of QAOA and allows an extension of typical cut fractions from the large $N$ limit where most subgraphs are trees to a small $N$ limit where typically a very small fraction of subgraphs are trees.

\begin{figure}
    \centering
    \includegraphics[width=\linewidth]{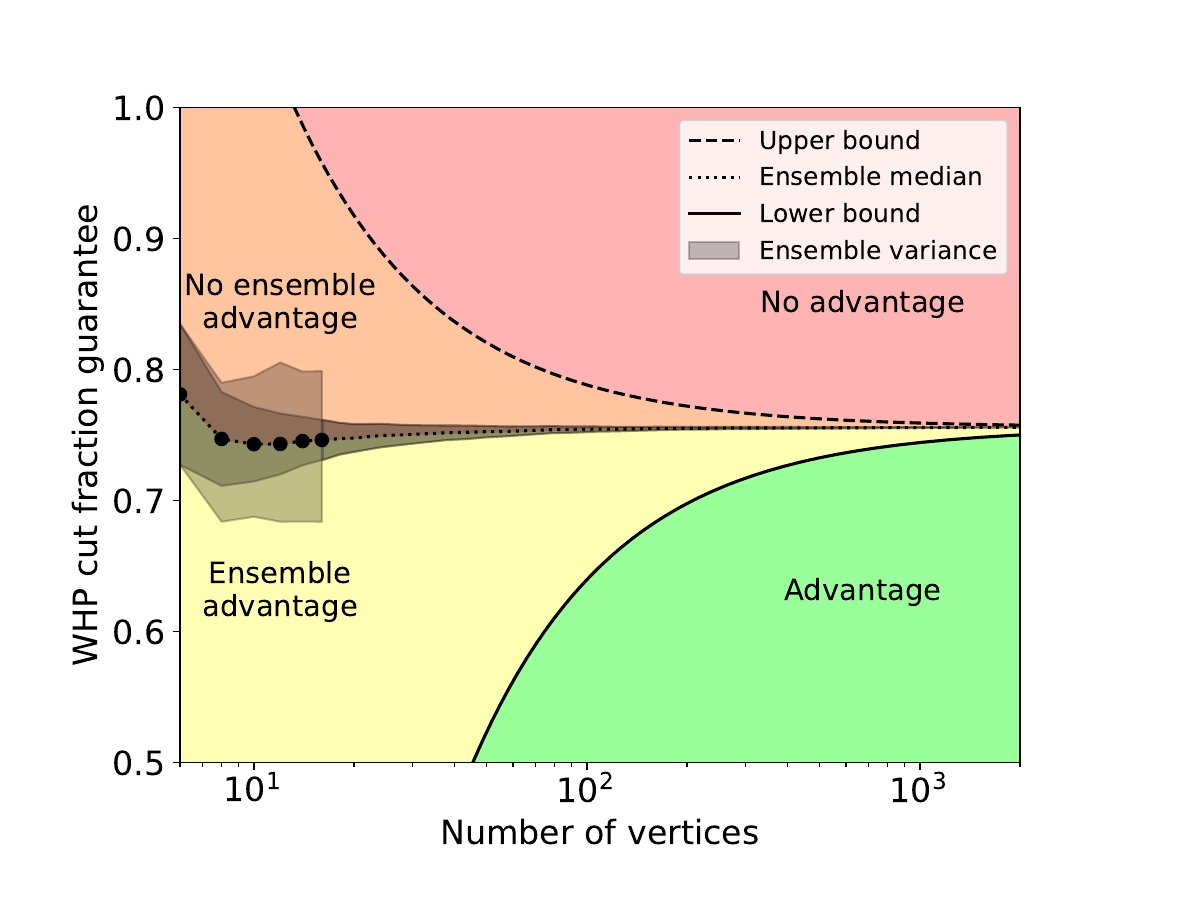}
    \caption{$p=2$ QAOA cut fraction guarantees under different assumptions. Dashed and solid lines plot with high probability the lower and upper bounds on cut fractions, respectively, assuming only graph theoretic typicality on the number of subgraphs. Dotted plots are the ensemble median over an ensemble of 3-regular graphs; for $N\leq 16$ (dots); this includes all graphs, while for $N>16$ this is an ensemble of 1,000 graphs for each size. We used 32 sizes between 16 and 256. Dark black fill plots the variance in the cut fraction over the ensemble, and light black fill plots the extremal values over the ensemble. The median serves as a proxy of performance assuming QAOA typicality. Given a particular cut from a classical solver, there may be different regions of advantage, shown by the four colors and discussed in the text.}
    \label{fig:bounds_p1}
\end{figure}

Figure \ref{fig:bounds_p1} plots the ensemble-averaged cut fraction for $p=2$ and various sizes of graphs. For $N\leq 16$, the ensemble includes every 3-regular graph (4,681 in total). For each size of $N>16$, we evaluate fixed-angle QAOA on 1,000 3-regular graphs drawn at random from the ensemble of all 3-regular graphs for each size $N\in (16,256]$. Note that because the evaluation is done at fixed angles, it may be done with minimal quantum calculation by a decomposition into subgraphs, then looking up the subgraph expectation value $f_\lambda$ from~\cite{Wurtz_guarantee}. This approach is also described in more detail in~\cite{shaydulin2021}. In this way, expectation values can be computed as fast as an isomorphism check.

From Fig.~\ref{fig:bounds_p1} we observe that the median cut fraction across the ensemble appears to concentrate around that of the tree subgraph value, even for ensembles where the typical graph is too small to include many tree subgraphs. Additionally, the variance (dark fill) reduces as $N$ increases, consistent with the fact that for larger $N$ there are fewer kinds of subgraphs with non-negligible frequency. Furthermore, the absolute range (light fill), which plots the largest and smallest expectation value across the ensemble, is consistently small. While the data for the absolute range  exists  here only for $N\leq 16$ because of complete sampling of the ensemble, 0ne can reasonably  expect that these absolute ranges extend for all $N$, suggesting that the absolute best performance of $p=2$ QAOA on 3-regular graphs is around $\approx 0.8$.

We numerically observe across a range of $p$ (not shown) that these behaviors persist: the typical cut fraction is approximately equal to that of the tree subgraph value $f_\text{p-tree}$ even in the limit where no subgraph is a tree. This suggests that the typical subgraph expectation value $f_\lambda\approx f_\text{p-tree}$, and only an atypical number of subgraphs have expectation values that diverge from the tree value. With this observation, we may use the value $f_\text{p-tree}$ as a proxy for the average cut fraction of fixed-angle QAOA.

\quad

These analyses yield four different regimes for advantage vs.~classical algorithms, shown in Fig.~\ref{fig:bounds_p1}. If a classical algorithm yields small cut fractions for large graphs (green, bottom right), then there is advantage in a strong sense. Based only on graph combinatorics, with high probability most of the edges participate in few cycles, and thus the cut fraction is almost guaranteed to be around the tree value, larger than the classical solver. Conversely, if the classical algorithm yields large cut fractions for large graphs (red, top right), there is no advantage in the strong sense: QAOA will  yield, for example, only $\sim 0.756$ for $p=2$ because most edges see no global structure. This analysis emphasizes that of \cite{farhi2020quantum}, which suggests that QAOA needs to `see' the whole graph in order to get reasonable performance.

Two additional performance regimes for small graphs exist, where QAOA can reasonably see the whole graph. If a classical algorithm yields small cut fractions for small graphs (yellow, bottom left), then there is advantage in a weak sense, which we call the `ensemble advantage'. Based on QAOA concentration, there is at least a $50\%$ chance that the QAOA result on a particular graph will yield a better cut fraction than will the classical algorithm; assuming that the variance in cut fraction is small, this is a `with high probability' statement. Conversely, if the classical algorithm yields large cut fractions for small graphs (orange, top left), there is no advantage in a weak sense. Assuming QAOA concentration, the cut fraction will be smaller than the classical value, and for some classical cut fraction there are no graphs with advantage (e.g., $>0.8$ for $p=2$).

Based on these numerical results, we may use the expectation value of the tree subgraph $f_\text{p-tree}$ as a high-probability proxy for typical fixed-angle QAOA performance on regular graphs. For large $N$, this result is validated by graph-theoretic bounds counting the typical number of tree subgraphs in a typical graph. For small $N$, this result is validated by fixed-angle QAOA evaluation on a large ensemble of graphs.

\begin{figure}
    \centering
    \includegraphics[width=\linewidth]{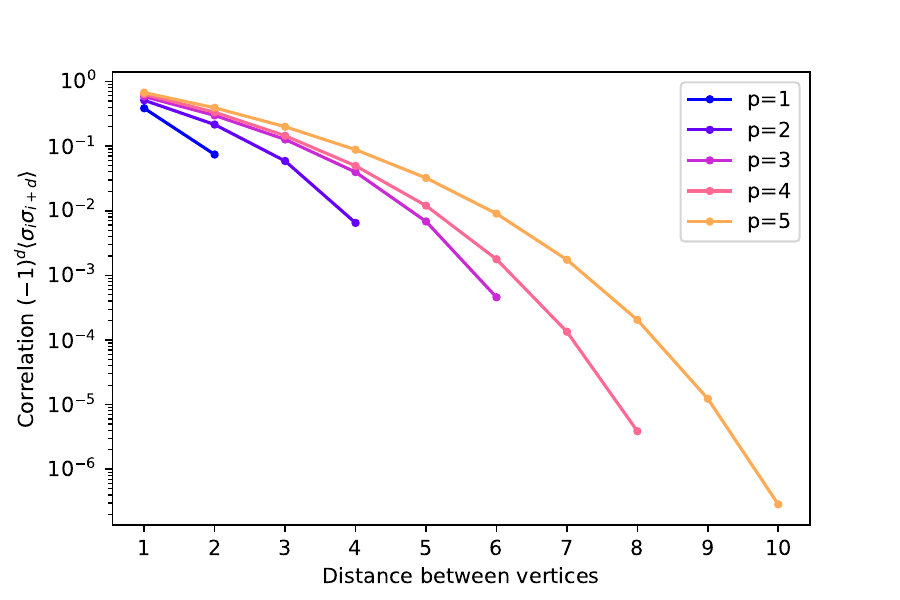}
    \caption{Long-range antiferromagnetic correlation coefficient on the 3-regular Bethe lattice, which is a proxy for an $N\to\infty$ typical 3-regular graph. Horizontal indexes the distance between two vertices.  QAOA is strictly local, which implies that  no correlations exist between vertices a distance $>2p$ away. As shown here, however, these correlations are exponentially decaying with distance. This suggests that even if the QAOA `sees the whole graph', one can use the central limit theorem to argue that the distribution of QAOA performance is  Gaussian with the standard deviation of $\propto 1/\sqrt N$}
    \label{fig:longrange_correlations}
\end{figure}

\subsection{Single-shot QAOA Sampling}
\label{sec:single-shot}

A crucial element of single-shot fixed-angle QAOA is  that the typical bitstring measured from the QAOA ansatz has a cut value similar to the average. This fact was originally observed by Farhi et al.~in the original QAOA proposal \cite{farhi2014quantum}: because of the strict locality of QAOA, vertices a distance more than $>2p$ steps from each other have a $ZZ$ correlation of strictly zero. Thus, for large graphs with a width $>2p$, by the central limit theorem the cut fraction concentrates to a Gaussian with a standard deviation of order $\frac{1}{\sqrt{N}}$ around the mean. As the variance grows sublinearly in $N$, the values concentrate at the mean, and thus with high probability measuring a single sample of QAOA will yield a solution with a cut value close to the average.

However, this result is limited in scope for larger depths $p$, because it imposes no requirements on the strength of correlations for vertices within distance $\leq2p$. Therefore, here we strengthen the argument of Farhi et al.~and show that these concentration results may persist even in the limit of large depth $p$ and small graphs $N$. We formalize these results by evaluating the $ZZ$ correlations of vertices within $2p$ steps, as shown in Fig.~\ref{fig:longrange_correlations}. Expectation values are computed on the 3-regular Bethe lattice, which has no cycles and thus can be considered the $N\to\infty$ typical limit. Instead of computing the nearest-neighbor correlation function, the x-axis computes the correlation function between vertices a certain distance apart. For distance 1, the correlations are that of the objective function $f_\text{p-tree}$. Additionally, for  distance $>2p$, the correlations are strictly zero in accordance with the strict locality of QAOA. For distance $\leq 2p$, the correlations are exponentially decaying with distance. Consequently, even for vertices within the lightcone of QAOA, the correlation is small; and so by the central limit theorem the distribution will be Gaussian.
This result holds because the probability of having a cycle of fixed size converges to 0 as $N\to \infty$. 
In other words, we know that with $N\to\infty$ we will have a Gaussian cost distribution with standard deviation $\propto\frac{1}{\sqrt N}$.

When considering small $N$ graphs, ones that have cycles of length $\leq 2p+1$, we can reasonably  extend the argument of Section~\ref{sec:ensemble_estimates} on typicality of subgraph expectation values. Under this typicality argument, the correlations between close vertices is still exponentially decaying with distance, even though the subgraph may not be a tree and there are multiple short paths between vertices. Thus, for all graphs, by the central limit theorem the distribution of solutions concentrates as a Gaussian with a standard deviation of order $\frac{1}{\sqrt{N}}$ around the mean. By extension, with probability $\sim 50\%$, any single measurement will yield a bitstring with a cut value greater than the average. These results of cut distributions have been found heuristically in \cite{larkin2020evaluation}.

\quad

The results are a full characterization of the fixed-angle single-shot QAOA on 3-regular graphs. Given a typical graph sampled from the ensemble of all regular graphs, the typical cut fraction from level $p$ QAOA will be about that of the expectation value of the $p$-tree $f_{\text{p-tree}}$. The distribution of bitstrings is concentrated as a Gaussian of subextensive variance around the mean, indicating that one can find a solution with quality greater than the mean with order 1 samples. Furthermore, because the fixed angles bypass the hybrid optimization loop, the number of queries to the quantum simulator is reduced by orders of magnitude, yielding solutions on potentially  millisecond timescales.

\subsection{Mult-shot QAOA Sampling}
\label{sec:multi-shot}

In the preceding section we demonstrated that the  standard deviation of MaxCut cost distribution falls as $1/\sqrt{N}$, which deems impractical the usage of multiple shots for large graphs. However, it is worth verifying more precisely  its effect  on the QAOA performance.
The multiple-shot QAOA involves measuring the bitstring from the same ansatz state and then picking the bitstring with the best cost. 
To evaluate such an approach, we need to find the expectation value for the best bitstring over $K$ measurements.

As shown above, the distribution of cost for each measured bitstring is  Gaussian, $p(x) = G(\frac{x-\mu_p}{\sigma_N})$.
We define a new random variable $\xi$ which is the cost of the best of $K$ bitstrings.
The cumulative distribution function (CDF) of the best of $K$ bitstrings is $F_K(\xi)$, and $F_1(\xi)$ is the CDF of a normal distribution.
The probability density for $\xi$ is

\begin{equation}
    p_K(\xi) = \frac{d}{d\xi} F_K(\xi) =\frac{d}{d\xi} F_1^K(\xi)
    = K F_1^{K-1}(\xi) p(\xi),
\end{equation}
where  $F_1(\xi) = \int_{-\infty}^\xi p(x) d x$ and $F_1^K$ is the ordinary exponentiation.
The expectation value for $\xi$ can be found by $E_K = \int_{-\infty}^\infty d x \, x p_K(x)$.
While the analytical expression for the integral can be extensive,  a good upper bound exists for it: $E_K\leq \sigma\sqrt{2 \log K} + \mu$.

Combined with the $1/\sqrt N$ scaling of the standard deviation, we can obtain a bound on
improvement in cut fraction from sampling $K$ times:

\begin{equation}
\label{eq:multi-shot}
    \Delta = \gamma_p\sqrt{\frac{2}{N} \log K},
\end{equation}
where $\gamma_p$ is a scaling parameter.
The value $\Delta$ is the difference of solution quality for multishot and single-shot QAOA. Essentially it determines the utility of using multishot QAOA.
We can determine the scaling constant $\gamma_p$ by classically simulating
the distribution of the cost value in the ansatz state. 
We perform these simulations using QTensor for an ensemble of graphs with $N\leq 26$ to obtain
$\gamma_6 = 0.1926$ and $\gamma_{11} = 0.1284$.

It is also worthwhile to verify the $1/\sqrt N$ scaling,
by calculating $\gamma_p$ for various $N$. We can do so for smaller $p=3$ and graph sizes $N\leq256$.
We calculate the standard deviation by $\Delta C= \sqrt{\langle C^2 \rangle - \langle C \rangle ^2}$ and evaluate the $\langle C^2 \rangle$ using QTensor. This evaluation gives large light cones for large $p$;  the largest that we were able to simulate is $p=3$.
From the deviations $\Delta C$ we can obtain values for $\gamma_3$. We find that for all $N$ the values stay within 5\% of the average over all $N$. This shows that they do not depend on $N$, which in turn signifies that the $1/\sqrt N$ scaling is a valid model. The results of numerical simulation of the standard deviation are discussed in more detail in the Supplementary Methods.

To compare multishot QAOA with classical solvers, we plot the expected performance of multishot QAOA in Fig.~\ref{fig:timebounds} as dash-dotted lines. We assume that a quantum device is able to sample at the 5kHz rate. Today's hardware is able to run up to $p=5$ and achieve the 5 kHz sampling rate \cite{Harrigan2021}.
Notably, the sampling frequency of modern quantum computers is bound not by gate duration, but by 
qubit preparation and measurement.

For small $N$, reasonable improvement can be achieved by using a few samples. For example, for $N=256$ with $p=6$ and just $K=200$ shots, QAOA can perform as well as single-shot $p=11$ QAOA. For large $N$, however, too many samples are required to obtain substantial improvement for multishot QAOA to be practical.

\subsection{Classical performance}\label{sec:classicalperformance}

To compare the QAOA algorithm with its classical counterparts, we choose the state-of-the art algorithms that solve the similar spectrum of problems as QAOA, and we  evaluate the time to solution and solution quality. Here, we compare  two algorithms: Gurobi and MQLib+BURER2002. 
Both are anytime heuristic algorithms that can provide an approximate solution at arbitrary time. For these algorithms we 
collect the `performance profiles'---the dependence of solution quality on time spent  finding the solution.
We also evaluate performance of a simple MaxCut algorithm FLIP. This algorithm has a proven linear time scaling with input size. It returns a single solution after a short time. To obtain a better FLIP solution, one may run the algorithm several times and take the best solution, similarly to the multishot QAOA.

Both algorithms have to read the input and perform some initialization step to output any solution. This initialization step determines the minimum time required for getting the initial 
solution---a `first guess' of the algorithm. This time is the leftmost point of the performance profile marked with a star in Fig.~\ref{fig:timebounds}. 
We call this time $t_0$ and the corresponding solution quality `zero-time~performance'.

We observe two important results.
\begin{enumerate}
    \item Zero-time performance is constant with $N$ and is comparable to that of $p=11$ QAOA,
    as shown in Fig.~\ref{fig:t0_cutf}, where solid lines show classical performance and dashed lines show QAOA performance.
    \item $t_0$ scales as a low-degree polynomial in $N$, as shown in Fig.~\ref{fig:adv_freq}. The y-axis is $t_0$ for several classical algorithms.
\end{enumerate}

Since the zero-time performance is slightly above the expected QAOA performance at $p=11$, we focus on analyzing this zero-time regime.
In the following subsections we discuss the performance of the classical algorithms and then proceed to the comparison with QAOA.

\subsection{Performance of Gurobi Solver}

In our classical experiments, as mentioned in Section~\ref{sec:meth_classical},
we collect the solution quality with respect to time for multiple $N$ and graph instances.
An example averaged solution quality evolution is shown in Fig.~\ref{fig:timebounds} for an ensemble of 256 vertex 3-regular graphs. Between times 0 and $t_{0, G}$, the Gurobi algorithm goes through some initialization and quickly finds some naive approximate solution. Next, the first incumbent solution is generated, which will be improved in further runtime. Notably, for the first 50 milliseconds, no significant improvement to solution quality is found. After that, the solution quality starts to rise and slowly converge to the optimal value of~$\sim 0.92$.

It is important to appreciate that Gurobi is more than just a heuristic solver: in addition to the incumbent solution, it always returns an upper bound on the optimal cost. 
When the upper bound and the cost for the incumbent solution match, the optimal solution is found.
It is likely that Gurobi spends a large portion of its runtime on proving the optimality by lowering the upper bound. This emphasizes that we use Gurobi as a worst-case classical solver.

Notably, the x-axis of Fig.~\ref{fig:timebounds} is logarithmic: the lower and upper bounds eventually converge after exponential time with a small prefactor, ending the program and yielding the exact solution. Additionally, the typical upper and lower bounds of the cut fraction of the best solution are close to 1. Even after approximately 10 seconds for a 256-vertex graph, the algorithm returns cut fractions with very high quality $\sim 0.92$, far better than intermediate-depth QAOA.

The zero-time performance of Gurobi for $N = 256$ corresponds to the Y-value of the star marker on Fig.~\ref{fig:timebounds}. We plot this value for various $N$ in Fig.~\ref{fig:t0_cutf}. As shown in the figure, zero-time performance goes up and reaches a constant value of $\sim 0.882$ at $N \sim 100$. Even for large graphs of $N = 10^5$, the solution quality stays at the same level.

Such solution quality is returned after time $t_{0,G}$, which we plot in Fig.~\ref{fig:adv_freq} for various $N$. For example, for a 1000-node graph it will take $\sim 40$ milliseconds to return the first solution. Evidently, this time scales as a low-degree polynomial with $N$. This shows that Gurobi can consistently return solutions of quality $\sim 0.882$ in polynomial time.

\subsection{ Performance of MQLib+BURER2002 and FLIP Algorithms}

The MQLib algorithm with the BURER2002 heuristic shows significantly better performance, which is expected since it is specific to MaxCut.
As shown in Fig.~\ref{fig:timebounds} for $N=256$ and in Fig.~\ref{fig:adv_freq} for various $N$, the speed of this algorithm is much better compared with Gurobi's. Moreover, 
$t_0$ for MQLib also scales as a low-degree polynomial, and for 1,000 nodes MQLib can return a solution in $2$ milliseconds.
The zero-time performance shows the same constant behavior, and the value of the constant is slightly higher than that of Gurobi, as shown in Fig.~\ref{fig:t0_cutf}.

While for Gurobi and MQLib we find the time scaling heuristically, the FLIP algorithm is known to have linear time scaling. With our implementation in Python, it shows speed comparable to that of MQLib and solution quality comparable to QAOA $p=6$.
We use this algorithm as a demonstration that a linear-time algorithm can give constant performance for large $N$, averaged over multiple graph instances.

\section{Acknowledgements}

This research was developed with funding from the Defense Advanced Research Projects Agency (DARPA). The views, opinions and/or findings expressed are those of the author and should not be interpreted as representing the official views or policies of the Department of Defense or the U.S. Government.
 Y.A.’s and D.L.'s work at Argonne National Laboratory was supported by the U.S. Department of Energy, Office of Science, under contract DE-AC02-06CH11357.
The work at UWM was also supported by the U.S. Department of Energy, Office of Science, National Quantum Information Science Research Centers.
\\

\section{Data availability}
The code, figures and datasets generated during the current study are available in a public repository
\url{https://github.com/danlkv/quantum-classical-time-maxcut}.
See the \texttt{README.md} file for the details on the contents of the repository.

\section{Author contribution}

D. L. and J. W. performed and analyzed the experiments and wrote the main text.
C. P. generated the FLIP data. M. S., T. N, and Y. A.
edited the paper.

\section{Competing interests}

T. N. and M. S. are equity holders of and employed by ColdQuanta, a quantum technology company.
J.W. is a small equity holder of and employed by QuEra Computing. The authors declare that the authors have no other competing interests

\bibliography{qis}

\pagebreak
\clearpage
\begin{widetext}
\begin{center}
\textbf{\large Supplementary Information for `Sampling Frequency Thresholds for Quantum Advantage of Quantum Approximate Optimization Algorithm'}
\end{center}
\end{widetext}
\setcounter{page}{1}
\setcounter{equation}{0}
\setcounter{figure}{0}
\setcounter{table}{0}
\makeatletter
\renewcommand{\figurename}{Supplementary Fig.}

\section*{Supplementary Methods}
\label{sec:supmeth}
\subsection{Quantum simulator QTensor}
\label{sec:qtensor}

The most popular method for quantum circuit simulation is state-vector evolution. It stores full state vector and hence requires memory size $\propto 2^N$, exponential in the number of qubits. It can simulate only circuits with $\lesssim 30$ qubits on a common computer, and simulation of 45 qubits on a supercomputer was reported \cite{statevector_sim}.
One can perform compression of the state vector and therefore simulate up to 61 qubits on a supercomputer \cite{Wu2019}.

For this work, we aim to study performance of QAOA on instances large enough to use in comparison with classical solvers. To simulate expectation values on lange graphs, we use the classical simulator \texttt{QTensor} \cite{qtensor, lykov2021large}. This simulator is based on tensor network contraction and allows for simulation of a much larger number of qubits. 
\texttt{QTensor} converts a quantum circuit to a tensor network, where each quantum gate is represented by a tensor. The indices of this tensor represent input and output subspaces of each qubit that the gate acts on. The tensor network constructed in this way can then be contracted in an efficient manner to compute the required value.
The contraction does not maintain any information about the structure of the original quantum circuit, and can result in significant simulation cost reduction, not limited to the QAOA context \cite{HenryML}. 

To calculate an expectation value of some observable $\hat R$ in a state generated by a circuit $\hat U$, one can 
evaluate the following expression: $\bra 0 \hat U^\dagger \hat R \hat U \ket 0$.
This value can be calculated by using a tensor network as well.
When applied to the MaxCut QAOA problem, the $\hat R$ operator is a sum of smaller terms, as shown in Eq. \ref{eq:maxcut_cost}.
The expectation value of the cost for the graph $G$ and QAOA depth $p$ is then

\begin{align*}
\braket{C}_p(\gb) &= \braket{\gb| \hat C | \gb} 
\\
&=  \sum_{\langle jk\rangle\in G}\langle\gb|\frac{1}{2}(1-\Z^j\Z^k)|\gb\rangle
\\
&=\sum_{\langle jk\rangle\in G}f_{\langle jk\rangle}(\gb),
\end{align*}
where $f_{\langle jk\rangle}=\langle\gb|(1-\Z^j\Z^k)|\gb\rangle/2$ is an individual edge contribution to the total cost function.
Each contribution to the cost function $f_{\langle jk\rangle}$ can be evaluated by using a corresponding tensor network.
Note that the observable $\Z^j\Z^k$ in the definition of $f_{\langle jk\rangle}$
acts only on two qubits and hence commutes with gates that 
act on other qubits.
The $\bra \gb$ state is not stored in memory at any time but rather is represented as a tensor network generated from the quantum circuit
shown in Eq.~\ref{eq:QAOA_ansatz}.
When two such tensor network representations (one for $\bra \gb$ and another for $\ket \gb$) are joined aside of the observable operator, it is possible
to cancel out the quantum gates that commute through the observable, thereby  significantly reducing the size of the tensor network.
The tensor network after the cancellation is equivalent to calculating $\Z^j\Z^k$ on a subgraph $S$ of the original graph $G$. 

While  multiple approaches exist for determining the best way to contract a tensor network, we use a contraction approach called bucket elimination \cite{detcher2013bucket}, which contracts one index at a time. At each step we choose some index $j$ from the tensor expression and then sum over a product of tensors that have $j$ in their index. The size of the intermediary tensor obtained as a result of this operation is very sensitive to the order in which indices are contracted.
To find a good contraction ordering, we use a line graph of the tensor network.
A tree decomposition~\cite{harvey2018treewidth} of the line graph corresponds to a contraction path that guarantees that the number of indices in the largest intermediary tensor will be equal to the width of the tree decomposition \cite{markov2008simulating}.
In this way one can simulate QAOA to reasonable depth on hundreds or thousands of qubits. More details of \texttt{QTensor} and tensor networks are in \cite{shutski2019adaptive, lykov2021large, Gray_Cotengra}.

\subsection{FLIP Algorithm}
\label{sec:flip}

As an example of a simple and fast classical MaxCut algorithm we evaluate a local search algorithm.
This class of heuristic is frequently referred to as FLIP \cite{favorablelandscapes} or FLIP-neighborhood \cite{elsaesser2011settling}.
A FLIP algorithm heuristic searches locally for improvements to the current solution that flip a single vertex. One still retains  the freedom to choose how to search for the vertex to flip at each stage of the algorithm. Examples of vertex selection methods include randomized message passing \cite{alaoui2021local} and zero temperature annealing.

The FLIP algorithm is as follows. First, for each vertex of a graph,  initially assign a value of 0 or 1 at random with equal probability. From this starting point, randomly order the vertices of the graph,  iterate through each vertex in order, and flip the vertex if flipping it will improve the cut value. Once all vertices have been iterated through, repeat this process until no vertices are changed in a full iteration. This procedure is analogous to  zero-temperature Monte Carlo annealing and is a greedy solver. The end result is a partition in which flipping any individual vertex will not improve the cut size. On 3-regular graphs we observe that this algorithm runs on graphs of $N=10,000$ nodes in about 70 ms on an Intel I9-10900K processor and gives a mean cut fraction of $0.847$,
which matches the performance of $p=6$ QAOA.

Given more time, the FLIP algorithm can improve its performance by reinitializing with a new random choice of vertex assignments and vertex orderings, as shown in Fig.~\ref{fig:timebounds}. Given an exponential number of repetitions, the algorithm will eventually converge on the exact result, although very slowly.

As a local binary algorithm, it runs into locality restrictions \cite{barak2021classical} and less-than-ideal performance but is extremely fast.
To put this into perspective with QAOA, we implemented FLIP using Python. We observe that a simple implementation returns solutions for a 100,000 vertex 3-regular graph in $<1$ second.
Optimized or parallelized implementation using high-performance languages such as C++ may run several times faster.
The main property is that for a graph of degree $k$, girth $L$, and size $N$, the FLIP algorithm runtime scales as $O(NLk)$ \cite{alaoui2021local},
which we verify experimentally. Notably, for any quantum initialisation step the time scaling would also be at least $O(N)$, since we have to somehow move information about the graph to the quantum device.

\subsection{Graph statistics bounds}
\label{sec:graph_stat_bounds}

\begin{figure*}
    \begin{center}
    \includegraphics[width=0.8\linewidth]{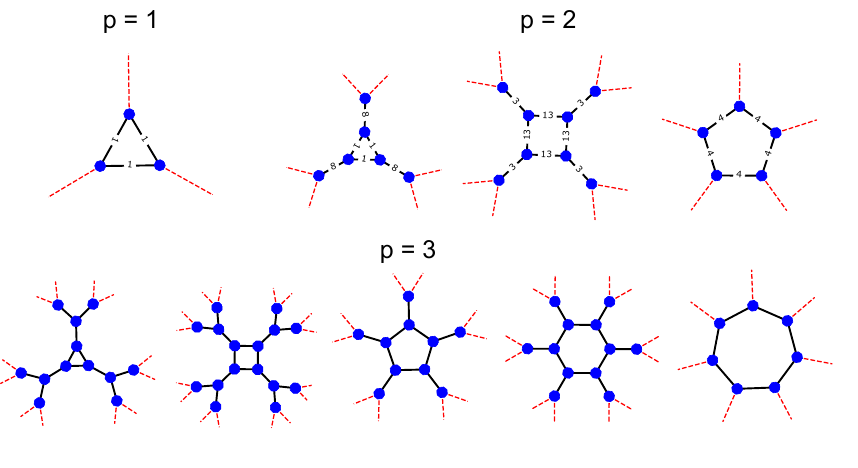}
    \end{center}
    \caption{Counting types of subgraphs on sparse cycles to find upper and lower limits on QAOA expectation values. The presence of a finite number of cycles in an infinitely large graph slightly modifies the value of the QAOA expectation value by modifying the local subgraphs. As shown above, the edges that are modified as a part of each cycle for $p\leq 3$ are shown in black; vertices which connect to the rest of the graph are shown in red. Edge labels refer to subgraph indexing in \cite{Wurtz_guarantee}.}
    \label{fig:cycle_subgraphs}
\end{figure*}

It is known \cite{WORMALD1981168,McKay2004} that in the limit $N\to\infty$, the probability of a graph having $t$ cycles of length $l$, where connectivity $d$ is odd, is asymptotic to

\begin{equation}
    P(l,t) = \frac{\lambda^t \exp(-\lambda)}{t!}\quad\text{where}\quad \lambda = \frac{(d-1)^l}{2l}.
\end{equation}

Summing over all $t$, we  find that the average number of cycles of size $l$ is equal to the same size-independent constant

\begin{equation}
    k_l = \sum_{t=0}^\infty tP(l,t) = \frac{(d-1)^l}{2l}.
\end{equation}

This is a probabilistic estimate based on statistics of regular graphs, does not depend on QAOA, and is asymtotically precise; thus, this is  a `with high probability' (WHP) result. For 3 regular graphs, the small values of $k_l= [ 1.33, 2.00, 3.20$, increasing exponentially. These sparse cycles modify the subgraph that QAOA `sees' in its local environment, as shown in Supplementary Fig.~\ref{fig:cycle_subgraphs}. Each cycle of length $l$ modifies the subgraph of

\begin{equation}
    m_l^p = l+l\sum_{q=0}^{p - \lceil l/2 \rceil}(d-1)^q
\end{equation}
edges; $l$ for the edges which participate in the cycle, plus some additional number that is exponential in $p$. For example, a 4-cycle of a 3 regular graph and $p=3$ modifies 16 edges, as shown in Supplementary Fig.~\ref{fig:cycle_subgraphs} This count of modified edges serves as an upper bound on the number of tree subgraphs

\begin{equation}\label{eq:nontree_lowerbound}
    M_\text{p-tree} \geq M - \sum_{l=3}^{2p+1} m_l^pk_l ,
\end{equation}
as an edge may participate in more than one cycle. Note that $M_\text{p-tree}$ should trivially be $\geq 0$, which occurs for large $p$ and small $N$. For $p=2$ and 3 regular graphs, this value is $M_\text{2-tree}\geq M - 40$, and serves as a characteristic scale for when QAOA will begin to see global structure.

The expectation value as a sum over subgraphs can then be broken into two parts: the tree subgraph and everything else. The sum can then be bounded by fixing an extremal value for the expectation value of every other subgraph, knowing that $0\leq f_\text{min}\leq  f_\lambda\leq f_\text{max}\leq 1$.

\begin{align}
    \langle \hat C \rangle &= M_\text{p-tree} f_\text{tree} + \sum_{\lambda\neq\text{p-tree}} M_\lambda f_\lambda,\\
    &\leq M - M_\text{p-tree}(f_\text{max}-f_\text{p-tree}),\\
    &\geq M_\text{p-tree}f_\text{p-tree}+(M-M_\text{p-tree})f_\text{min}
\end{align}

Combined with the lower bound of Supplemental Eq.~\ref{eq:nontree_lowerbound}, these bounds are
\begin{align}
    \frac{\langle \hat C\rangle}{M}\;\geq\;&f_\text{tree} - \sum_{l=3}^{2p+1} \frac{m_l^pk_l}{M}(f_\text{min} - f_\text{tree}),\\
    \frac{\langle \hat C\rangle}{M}\;\leq\;&f_\text{tree} + \sum_{l=3}^{2p+1} \frac{m_l^pk_l}{M}(f_\text{max} - f_\text{tree}).
\end{align}

Using the enumeration of subgraphs from \cite{Wurtz_guarantee}, $f_\text{min,p=2}=0.4257$ and $f_\text{max,p=2}=0.8771$. Thus, for $p=2$ and 3 regular QAOA, the performance can be bounded to be between

\begin{equation}
    0.7559 - \frac{13.208}{M} \leq \frac{\langle \hat C\rangle}{M} \leq 0.7559 + \frac{4.848}{M}.
\end{equation}

Therefore, for $N$ large, the value of $\langle\hat C\rangle$ is bounded from above and below by a constant amount and converges to the tree value. Similarly, for $N$ small, the value of $\langle \hat C\rangle$ is bounded between $0$ and $1$ as WHP every edge participates in at least one cycle of length $l\leq2p+1$ and so there are no tree subgraphs to contribute to the count. In principle, these bounds may be tightened by including expectation values $f_\lambda$ for more subgraphs.

This is a `with high probability' result: there may be extremely atypical graphs that have much different numbers of tree and single cycle subgraphs. For example, an atypical graph may be one of size $N$ that is two graphs of size $N/2$ connected by a single edge. This bound is a generalization of the work of \cite{farhi2020quantum}, which observes that the QAOA needs to `see the whole graph' in order to have advantage. Here the upper and lower bounds are based on the same argument, except generalized to the small-$N$ regime

\subsection{Experimental validation of standard deviation scaling}
\label{sec:sd_experiment}

We verify numerically the $\propto \frac{1}{\sqrt N}$ scaling of standard deviation using two approaches.
Besides the validation of the theoretical results, these calculations allow to find the scaling coefficient to use in Equation~\ref{eq:multi-shot}.

The first approach is to calculate multiple probability amplitudes for each graph and estimate the variation by drawing a large number of samples from the distribution. This approach is feasible for graphs of small size and large $p$ since it only requires to calculate a subset probability amplitudes. For size $N<30$ any $p$ is feasible since it's possible to fit the full statevector in memory.
The second approach is to construct tensor network for observable $\langle C^2 \rangle$, then calculate the variance using $V = \langle C^2 \rangle - \langle C \rangle^2$. The quadratic observable has to be calculated for each pair of edges, and introduces $\propto N^2$ tensor networks. This approach allows to get the exact value of standard deviation without sampling error. Additionally, it allows to apply the lightcone optimization, which reduces computational cost for large graphs. However, the complexity grows rapidly with $p$ and we observe that only $p=2, 3$ are feasible on our hardware.

\begin{figure}
    \centering
    \includegraphics[width=\linewidth]{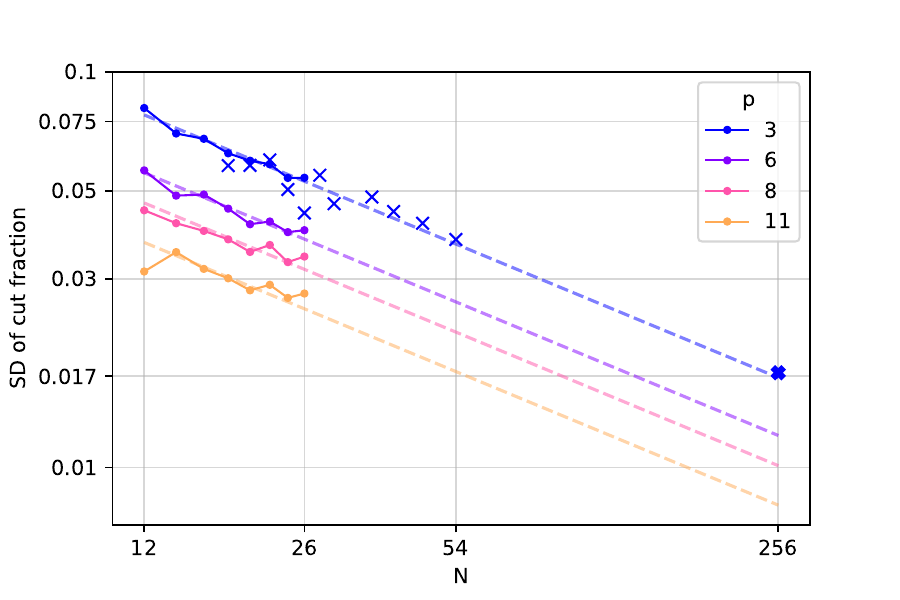}
    \caption{Dependence of standard deviation of MaxCut cut fraction on $N$ and $p$ for random 3-regular graphs. 
    Circle markers represent approximate evaluations over 1000 samples per graph and 20 graphs for each size $N$. The dashed line shows a fit to the approximate data using the $\propto \frac{1}{\sqrt N}$ scaling.
    Cross markers show exact standard deviation values for larger $N$ and $p=3$ with one graph per each $N$. These values were obtained using tensor network contraction via QTensor. 
    The bold cross marker at $N=256$ is also an exact value of QAOA standard deviation at size corresponding to $N$ used on Figure~\ref{fig:timebounds}.
    Note that this plot has a log-log scale.
    }
    \label{fig:sd_vs_n}
\end{figure}

We can see that for $N\leq26$ there is good agreement with theoretical predictions across different values of $p$.
We use these approximate small-$N$ results to find the scaling coefficient for each $p$. This will
be used in the rest of the paper to estimate the standard deviation for arbitrary $N$ and fixed $p$. In particular, Figure~\ref{fig:adv_rate} uses these estimations to obtain number of multi-shot QAOA samples.

To further verify the validity of our prediction, we use exact calculations of variance via QTensor.
These values are shown as crosses on Supplementary Fig.~\ref{fig:sd_vs_n} and are evaluated only for one graph for each graph size $N$.
For small $N$ these results are susceptible to the error due to using a single graph, since at smaller size graphs have diverse lightcones. However, as we see at larger $N$ and $N=256$ in particular, there is a good agreement 
with the predicted value of the standard deviation.

\subsection{Optimal time to compete with classical algorithms}
\label{sec:opt_time}

In the main text we were comparing zero-time performance of classical algorithms and QAOA.
This is justified since the classical algorithms usually improve their solution quality faster with time than multi-shot QAOA. However, this can be not the case for intermediate $N$ where the standard deviation of the cut fraction is relatively large. This means that using a time after $t_{0, G}$ might be better for QAOA. 
We study this possibility by finding the sampling frequency for each point $(\Delta(t), t)$ on the performance profile for each graph. The minimum sampling frequency would match at least one point on the performance profile and thus would be sufficient for performance advantage, albeit in a very narrow regime.

\begin{figure}[h]
    \includegraphics[width=\linewidth]{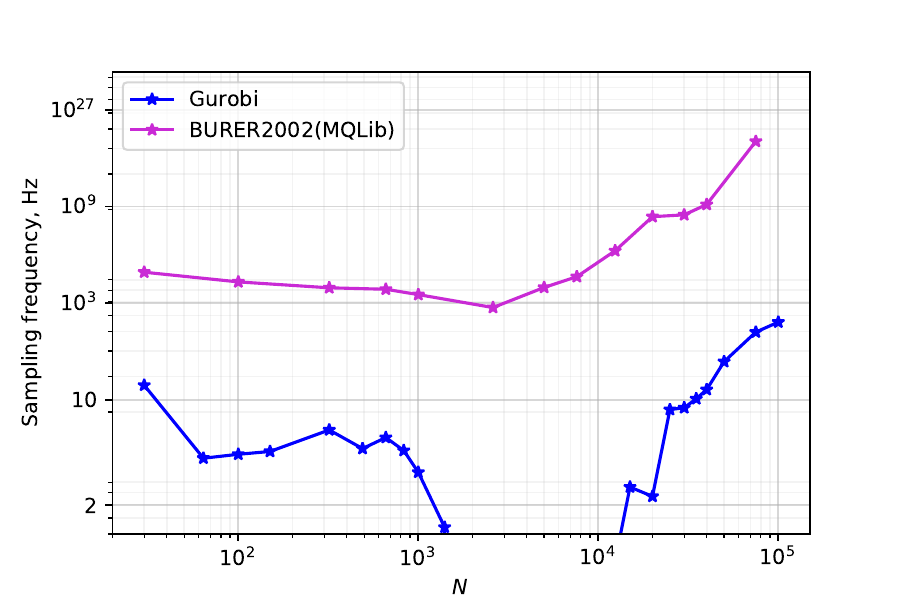}
    
    \caption{Optimal sampling frequency considering across all performance profile for classical algorithms
    The value is averaged over 100 seeds for Gurobi and 20 seeds for BURER2002.
    }
    \label{fig:opt_freqs}
    
\end{figure}

We plot the average minimum sampling frequency on Supplementary Figure~\ref{fig:opt_freqs}. 
We find that for the BURER2002 algorithm, there is no change from Figure~\ref{fig:adv_rate}, since this algorithm 
always scales better than the multi-shot QAOA. On the other hand, apparently it is enough to have a kHz sampling rate to 
match Gurobi before it starts to improve its solution. 
Note that the frequency for matching Gurobi can be rather noisy, since it is very sensitive to $\Delta$. From Fig.~\ref{fig:t0_cutf} one can see that Gurobi's solution quality in region $N=1000\dots10,000$ is smaller than that of QAOA $p=11$, a small change which could be caused by chance. This, however, results in a significantly small sampling frequency value on Supplementary Fig.~\ref{fig:opt_freqs} in the same range of $N$.


\end{document}